\documentclass[12pt]{article}
\usepackage[dvips]{color}
\usepackage{epsfig}
\usepackage{amsmath}
\usepackage{graphicx}

\begin{document}

\title{\bf Cosmic String in the BTZ Black Hole Background with Time-Dependant Tension}
\author{{ J. Sadeghi $^{a}$\thanks{Email: pouriya@ipm.ir}\hspace{1mm},
H. Farahani$^{a}$ \thanks{Email: h.farahani@umz.ac.ir}}, B. Pourhassan$^{a}$ \thanks{Email: b.pourhassan@umz.ac.ir} and  S. M. Noorbakhsh$^{a}$\thanks{Email: s.m.noorbakhsh@umz.ac.ir} \\
$^a${\small {\em Sciences Faculty, Department of Physics, Mazandaran
University,}}\\{\small {\em P .O .Box 47415-416, Babolsar, Iran}}}

\maketitle

\begin{abstract}
In this paper, we study the equation of circular loops  with
time-dependant tension in the BTZ black hole background. We obtain
various cases where cosmic string loops finally collapse to form
black holes. Also, we study effect of the BTZ black hole mass and
angular momentum on the evolution of cosmic string loops. We find
the critical values of initial radii as a limit for the cosmic
string loops collapsing to form black holes.\\\\
{\bf Keywords:} BTZ black hole; Time-dependant tension; Cosmic
string loops.
\end{abstract}
\section{Introduction}
As we know the interaction between two fields plays an important
role in cosmology and particle physics [1]. The connecting link was
provided by Kirzhnits [2] who suggested that spontaneously broken
symmetries can be restored at sufficiently high temperatures.
According to the modern ideas, the elementary particle interactions
are described by a grand unified theory (GUT) with a simple gauge
group $G$ which is a valid symmetry at  highest energies. As the
energy is lowered, the model undergoes a series of spontaneous
symmetry breaking [1]. On the other hand in hot big bang cosmology
this spontaneous symmetry breaking implies a sequence of phase
transitions [3]. The phase transitions in the early universe can
give rise to topological stable defects like strings. A discussion
of their evolution were given by Kibble [4]. Cosmic strings may be
created as one-dimensional topological space-time defects and are
hypothesized to form when the field undergoes a phase change in
different regions of space-time. This is  resulting in condensations
of energy density at the boundaries between regions. These cosmic
strings can lead to very interesting cosmological consequences such
as explaining galaxy formation [5-7]. Because of their space-time
metric with a deficit angle they can produce a number of distinctive
and unique observational effects. Also  Cosmic strings can be
produced at the end of inflation or at the end of brain inflation
[8,9]  providing us with a potential window on $M$-theory [10-12].
However, the CMB rules out cosmic strings as seeds of large scale
structure formation in the universe because of limits on their
tension $(G\mu < 10^{-6})$ [13-15]. Furthermore, cosmic strings
still have strong influence on  astrophysics [16, 17] such as
gravitational lensing effects [18, 19], gravitational wave
background [20, 21] and early re-ionization [22, 23]. They would
have immense density  and also would represent significant
gravitational sources. A cosmic string with 1.6 kilometers in length
would exert more gravity than the Earth. Cosmic strings would form a
network of loops in the early universe and their gravity could have
been responsible for the original clumping of matter into galactic
superclusters. After the formation of cosmic string, they are not
static and would continuously evolve under the force of their
tension. They can collide and intersect to undergo reconnections
also strings stretch under the influence of the Hubble expansion or
the environment. In that case  the strings lose energy to
gravitational radiation when they oscillate. The reconnections of
long strings and large loops will produce small loops copiously
[24]. The observational results to support the existence of cosmic
string loops in our Universe are in the Refs. [19, 25]. A cosmic
string's vibration, which would oscillate near the speed of light,
can cause parts of the string to pinch off into an isolated loop.
These loops have a finite lifetime due to decay via gravitational
radiation. Gravitational lensing of a galaxy by a straight section
of a cosmic string would produce two identical undistorted images of
the galaxy. A gravitational lens called CSL-1 which invokes two
images with comparable magnitudes of the same giant elliptical
galaxy was discovered. It is interesting that many similar objects
were found in Refs. [18, 26]. In 2003, a group led by Mikhail Sazhin
reported the accidental discovery of two seemingly identical
galaxies very close together in the sky, leading to the speculation
that a cosmic string had been found [27]. However, observations by
the Hubble Space Telescope in January 2005 showed them to be a pair
of similar galaxies, not two images of the same galaxy [28, 29]. A
cosmic string would produce a similar duplicate image of
fluctuations in the cosmic microwave background, which might be
detectable by the upcoming Planck Surveyor mission. As a complicated
time dependent gravitational source, the cosmic string loops
oscillate with time rather randomly. Schild et al observed the
brightness fluctuation in the multiple-image lens system Q0957+561A,
B [19, 25]. They think that system consists of two quasar images
separated by approximately 6 degrees. The phenomena are images of
the same quasar because of the same spectroscopic , fluctuating  in
brightness and time delay between fluctuation sit which is due to
lensing that they supply this synchronous variations in two images.
The cosmic string loops can also generate more gravitational waves
and distinct signatures [20, 21, 30], nearly all loops will become
black holes except in special cases. In Minkowski space-time and
Robertson-Walker universe, the loops will collapse to form black
holes under their own tension certainly instead of remaining
oscillating loops after the loops formation [31, 32]. In de Sitter
backgrounds, only loops with large initial radii can avoid becoming
black holes [31, 33, 34]. However a large loop will evolve to be a
lot of smaller loops, loops can not live in a de Sitter space-time
unless they are very large. Clearly, in the environment with a
positive cosmological constant, few cosmic string loops can survive.
The black holes are final results for loops of cosmic string if the
tension of cosmic string is constant [35]. So far in nearly all
researches considered the tensions of cosmic strings to be constant
which is just an assumption but M. Yamaguchi put forward the
important issue that the tensions of cosmic strings can depend on
the cosmic time [36, 37]. As we know in context of cosmological
phase transitions, when the field theory under symmetry group  $G$
as $U(1)$ is broken, the cosmic strings are formed.  So, we consider
a Higgs field $\phi$ with  following  self-interaction  potential,
\begin{equation}\label{E1}
V(\phi)=\frac{1}{2}\lambda(\phi^\dagger\phi -\eta^{2}),
\end{equation}
where $\phi$ is a complex field and $ \lambda$ is positive integer,
also $\mu$ is a constant. When symmetry is broken the vacuum
expectation value of $\phi$ is $<\phi>=\eta$ and the tension of
cosmic strings $\mu$ is given by $\mu\simeq \eta^2$. In the high
temperature limit we have an additional potential,
\begin{equation}\label{E1}
V_T(\phi)=AT^2(\phi^\dagger\phi+V(\phi)).
\end{equation}
As we see the effective mass of the field $\phi\propto \eta^2$ thus
$\eta$ depends on the temperature $T$ and hence on the cosmic time
$\eta$  is $\eta=\eta(t)$ [1]. This implies that the tension $\mu$
also depends on the cosmic time $\mu(t)$ [36]. Since $\eta$ is
associated with scale factor $a(t)$ then the tension can be denoted
as $\mu\propto a(t)^{-3}$ where $a(t)$ is the scale factor. As we
know $a(t)\propto t^{\frac{2}{3}}$, in a matter-dominated universe,
and also $a(t)\propto t^{\frac{1}{2}}$, in a radiation-dominated
universe. So the tension become proportional to $t^{\frac{2}{3}}$
and $ t^{\frac{1}{2}}$ respectively. We have a general case of
power-law expansion: $a(t)\propto t^{\alpha}$ [1]. The results show
that time varying tensions of cosmic strings can cause matter power
spectra and CMB completely different from those induced by constant
tensions [36, 37].\\
In order to explain cosmic string dynamic we need to find the
Nambu-Goto action. In that case we choose $\xi^0 ,\xi^1 $ parameters
as time like and space like coordinates respectively. If we choose
the gauge condition on string world-sheet coordinates we may use the
conformal gauge where $\dot x^2+x^{\prime 2}=1$, so we will obtain
wave equations as $\ddot{x}^\nu+x^{\prime\prime \nu}=0 $ or
$\ddot{x}^\nu-x^{\prime\prime\nu}=0 $ [38, 39]. In this case the
motion of the string must be periodic in time with the period,
$T=m/2\mu$ where $m$ is the mass of the loop [40]. In this paper, we
consider the wave equation as $\ddot{x}^\nu-x^{\prime\prime
\nu}=0 $.\\
All above information give us motivation to consider circular loop
with time dependent tension, and study the equation of cosmic string
loops in Banados , Teitelboim and Zanelli (BTZ) black hole
background. Also we investigate several cases of the BTZ black hole
with circular loop information. The idea that cosmic string loop
will form a black hole is not new. This subject already investigated
[41-44]. Now, we study such phenomena in BTZ background. Throughout
this paper we use natural units where the speed of light $c=1$.

\section{The BTZ black hole metric}
As we know the black hole solution of BTZ in (2+1) space time are
derived by the following  three dimensional theory of gravity,
\begin{equation}\label{E1}
S=\int d^{3}x \sqrt{-g}(^{(3)}R +\Lambda),
\end{equation}
where $^{(3)}R$ is called Rici scalar in three dimensions, and
$\Lambda$ denotes cosmological constant. The BTZ black hole is
obtained as the quotient space $SL(2,R) < (\rho_l , \rho_R)>$, where
$ < (\rho_l , \rho_R)>$ denote the subgroup of $SL(2,R)\times
SL(2,R)$ generated by $(\rho_l,\rho_R)$.
 In the coordinate chosen here the metric reads
\begin{equation}\label{E1}
dS^2=-\left(-M+\frac{r^2}{l^2}+\frac{J^2}{4r^2}\right)dt^2+\frac{dr^2}{(-M+\frac{r^2}{l^2}+\frac{J^2}{4r^2})}+r^2(d\phi
-\frac{J}{2r^2}dt)^2,
\end{equation}
where $M$ is Arnowwitt-Deser-Misner (ADM) mass , and $J$ is angular
momentum (spin) of the BTZ black hole where $-\infty < t <\infty$,
$0\leq R <\infty $ and $0<\theta < 2\pi$ with the cosmological
constant $\Lambda =\frac{1}{l^2}>0$. The metric (4) is singular when
$r=r_{\pm}$,
\begin{equation}\label{E1}
r_\pm ^2=\frac{Ml^2}{2}(1\pm[1-(\frac{J}{Ml})^2]^\frac{1}{2}),
\end{equation}
and,
\begin{equation}\label{E1}
M=\frac{r_+^2 + r_-^2}{l^2} , J=\frac{2r_+r_-}{l}.
\end{equation}
Here, we emphasis that if $|J|>Ml$, $r_{\pm}$ become complex and the
horizon  will disappear. In the case of $M=1$ and $J=0$ the metric
may be recognized as the ordinary AdS space.

\section{The circular loop equation}
Now, we study behavior of a circular cosmic string in the BTZ black
hole background (4). A free string propagating in a space-time
sweeps out a two-dimensional surface which is called string
world-sheet. The Nambu-Goto action for a cosmic string with
time-dependent tension is given by
\begin{equation}\label{E1}
S=-\int d^{2}\sigma \mu(\tau)\left[(\frac{\partial x}{\partial
\sigma ^0} \frac{\partial x}{\partial \sigma ^1}) ^2 -
(\frac{\partial x}{\partial \sigma ^0})^2 (\frac{\partial
x}{\partial \sigma ^1})^2\right]^{\frac{1}{2}},
\end{equation}
where $\mu(t)$ is the string tension which is  function of cosmic
time. Here, $\sigma ^a =(t,\phi)(a=0,1)$ are timelike and spacelike
string coordinates respectively. $x^\mu(t,\phi)$ with $\mu,\nu
=0,1,2,3$ are the coordinates of the string world-sheet in the
space-time. For simplicity and without generality changed we assume
that the string lies in the hypersurface $\theta=\frac{\pi}{2}$,
then the space-time coordinates of the world-sheet parameterized by
$\sigma^0=t$ ,$\sigma^3=\phi$, so we have
$x(t,r(t,\phi),\frac{\pi}{2}, \phi)$ as string profile. In the case
of planar circular loops, we have $r = r(t)$. According to the
metric (4) and the $d$-dimensional space-time coordinates mentioned
above, the Nambu-Goto action with an additional factor for the
time-dependent tension belonging to a cosmic string denoted as,
\begin{equation}\label{E1}
S=-\int dt d\phi \mu(t)r(h-\frac{\dot{r}^2}{h})^{\frac{1}{2}},
\end{equation}
where $h=(-M+\frac{r^2}{l^2}+\frac{J^2}{4r^2})$. Let $l=1$, then we
introduce the following equation for loops, which have time-varying
tension as $\mu(t)=\mu_0t^q$,
\begin{equation}\label{E1}
h^2\ddot{r}r+\frac{q}{t}r\dot{r}(h^2-\dot{r}^2)-\dot{r}^2(h^2+3r^2h)+h^4+r^2h^3-\frac{J^2}{4r^2}(h^3-3\dot{r}^2)=0,
\end{equation}\label{E3}
where $\mu_0$ is a constant. Recently people have solved the
equations for circular loops of cosmic string  for the Minkowski
space-time, Robertson-Walker universe, de-Sitter space-time  and
Kerr-de Sitter space-time [16, 32, 33, 35, 45].  They have found
some constraints that a loop of cosmic string should not collapse to
form black hole for the case of constant and time-dependent tension.
Time-dependent tension changes as the power $q$ like
$\mu(t)=\mu_0t^q$ in different space-times.  For example, Vilenkin
in the Ref. [16] shown that in case of Minkowski and
Friedmann-Robertson-Walker (FRW) space-time with constant tension
for cosmic strings, all closed loops finally will be collapsed to
form black hole. Then, Cheng and Liu [30] studied equation of
circular loops of cosmic string with time-dependent tension in the
Minkowski and FRW space-time and shown that there must exist a
critical value $q_f = -0.131.$ In the Minkowski space-time when $q <
q_f$  all cosmic string loops will expand to evolve or contrarily
will collapse to form black holes when $q > q_f$. For FRW background
it is shown that there also exist a special value for every era,
$q_r = -0.078$ for radiation-dominated era and $q_m = -0.062$ for
matter-dominated one. In each era when $q < q_r$ or $q < q_m$  all
cosmic string loops will expand to evolve with any values of $r(t_0)
= r_0$ as the initial radii. So, Larsen [33] solved equation of
loops in de-Sitter space-time with constant tension, it was shown
that loops of cosmic string will keep on expanding in de Sitter
space-times if their radii satisfies the condition like $r(t_0)>
0.707$. After that Cheng and Liu [35] solved equation of loops in
de-sitter space-time with time-dependent tension. They have shown
that  the results  same as Larsen [33] when the initial radius is
$r(t_0) > 0.707$. But in the case of $r(t_0) < 0.707$ there must
exist a critical value denoted as $\alpha$, when $q < \alpha$, the
cosmic string loops will enlarge to evolve or contrarily will
collapse to form black holes when $q
>\alpha$. Now, we are going to solve equation (9) numerically, for the case of
$q=0$ and $q\neq 0$. In that case we account $M$ and $J$ as
variables, we find constraint for initial radii of cosmic string
loops collapsing to form black holes or vice versa will expand to
evolve. Our numerical method is as the following. Once, we fix $q$,
$M$ and $J$, and vary initial radii to obtain curves of $r$ in terms
of cosmic time $t$. It help us to find critical values of initial
radii. Then, we fix $q$, $J$ and initial radii, and vary $M$ to
obtain a relation between the black hole mass and collapsing of the
loops to black hole. Finally we fix $M$, $J$ and initial radii, and
vary $q$ to obtain special region of $q$ where loops forming black
hole. First, we suppose tension as constant $q=0$, $J=0$ and $M=1$,
we get the same result as de-Sitter space-time solved by Larsen
[33]. In that case for collapsing to form black holes we have
$r(t_0)<0.7070$. In the case of $J\neq0$ and $M=1$, the results are
different and they are classified in the table 1.

\begin{center}
  \begin{tabular}{|c||c|}
    \hline
    $J$ &  $Critical        $ $r(t_0)$ \\ \hline\hline
    -3 & $r(t_0)>4.80$ \\ \hline
    -2 & $r(t_0)>3.740$ \\ \hline
    -1 & $r(t_0)>0.707$ \\ \hline
    $-2/3$ & $0.357<r(t_0)<0.707$ \\ \hline
    -1/2  & $0.26<r(t_0)<0.707$ \\ \hline
    0 & $r(t_0)<0.707$ \\ \hline
    1/2 & $0.26<r(t_0)<0.707$ \\ \hline
    2/3 & $0.357<r(t_0)<0.707$ \\ \hline
    1 & $r(t_0)>0.707$\\ \hline
    2 & $r(t_0)>3.740$ \\ \hline
    3 & $r(t_0)>4.80$ \\ \hline
    \hline
  \end{tabular}
\end{center}
\hspace{4.50cm} Table 1: $M=1, q=0$ and different $J$.\\

In the table 1. we present some critical initial radii of the loops
which lead to form black holes. It is completely obvious, in the
case of $|J|\geq1$, it should be $r(t_0)>\alpha$ where $\alpha$ is
the critical value of radii, we see that by increasing positive and
negative $|J|$ the critical initial radii become larger and we see
same symmetry in values of $J$ and radii. If $|J|<1$, then not only
we satisfy the above information but also there is an extra
constraint for initial radii as $\alpha<r(t_0)<0.707$ ($r(t_0)$ is
the initial radii).

\begin{figure}[th]
\begin{center}
\includegraphics[scale=.39]{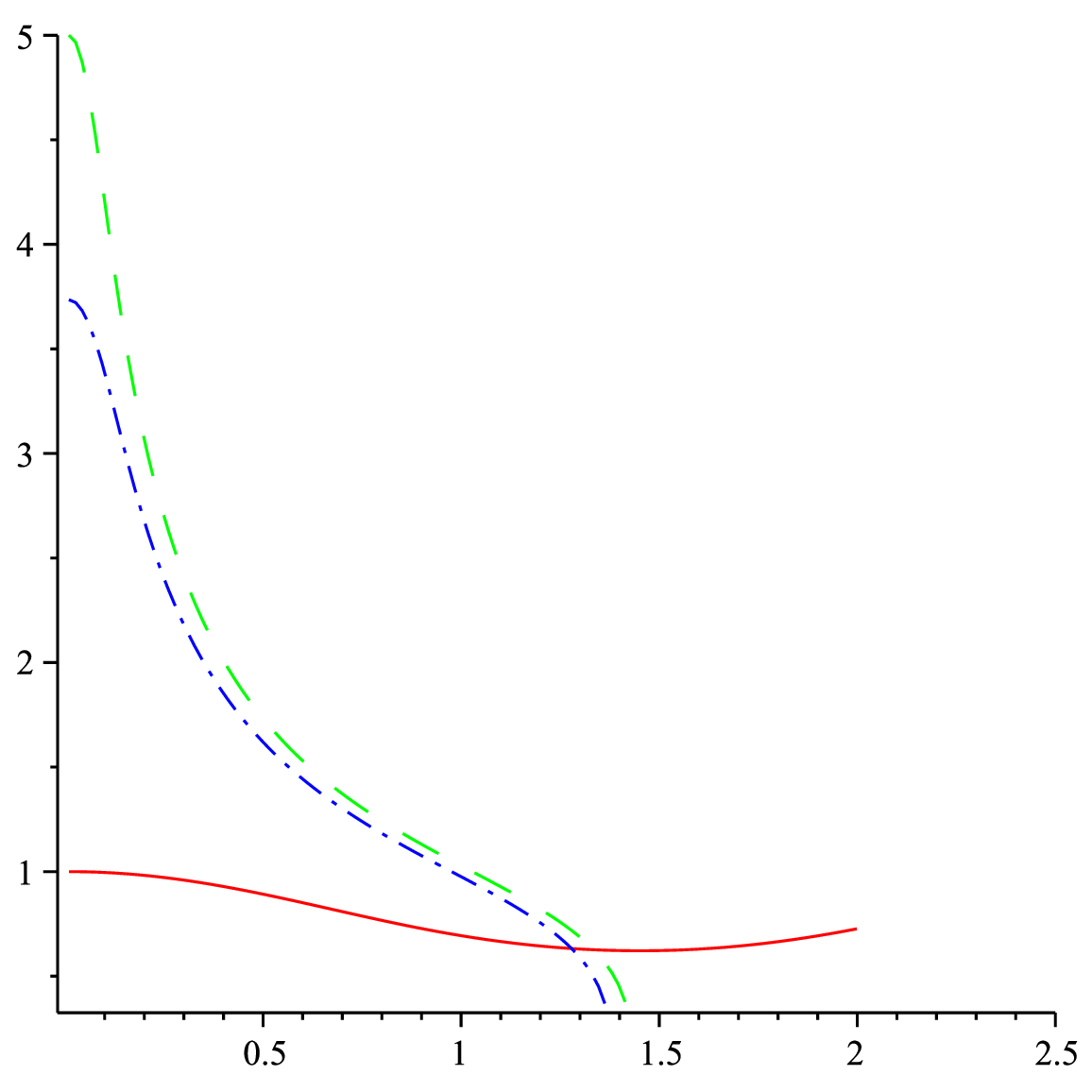}
\caption{Graph of the $r(t)$, radii of circular loops, in the BTZ
black hole with $M=1$ and $J=2$. The solid, dash-doted and dashed
curves drawn with initial value $r(t_0)$ =1.00 ,3.740 and 5.00
respectively under assumption of $\dot{r}=0$ and $q=0$.}
\end{center}
\end{figure}

We see in the table 1. results for $|J|>M$ and $|J|<M$ are
different. The case of $|J|>M$, $(M=1)$ implies that how much the
angular momentum of the black hole increases, only cosmic loops of
larger radii can form black holes and all the cosmic loops having
smaller radii than critical initial radii will survive. Since
$0.707$ is a very big amount then, most of the loops will keep on
expanding and the black
hole formation possibility will decline.\\
We see in the Fig. 1,
with $|J|>1$ the critical value of initial radii is larger than
$l=1$ which means that $r(t_0)$ is larger than
the size of the universe so their existence in such condition may be impossible.\\
But we see in the Fig. 2, for the $|J|<M$, with $M=1$, the result is
completely different. Here, loops of larger than $0.707$ will remain
in our universe and smaller ones will collapse to form black holes
then, there will be more possibility of forming black holes.\\
For other values of mass as a constant and different $J$ results are
the same and the only difference is in their critical initial
radii.\\

\begin{figure}[th]
\begin{center}
\includegraphics[scale=.4]{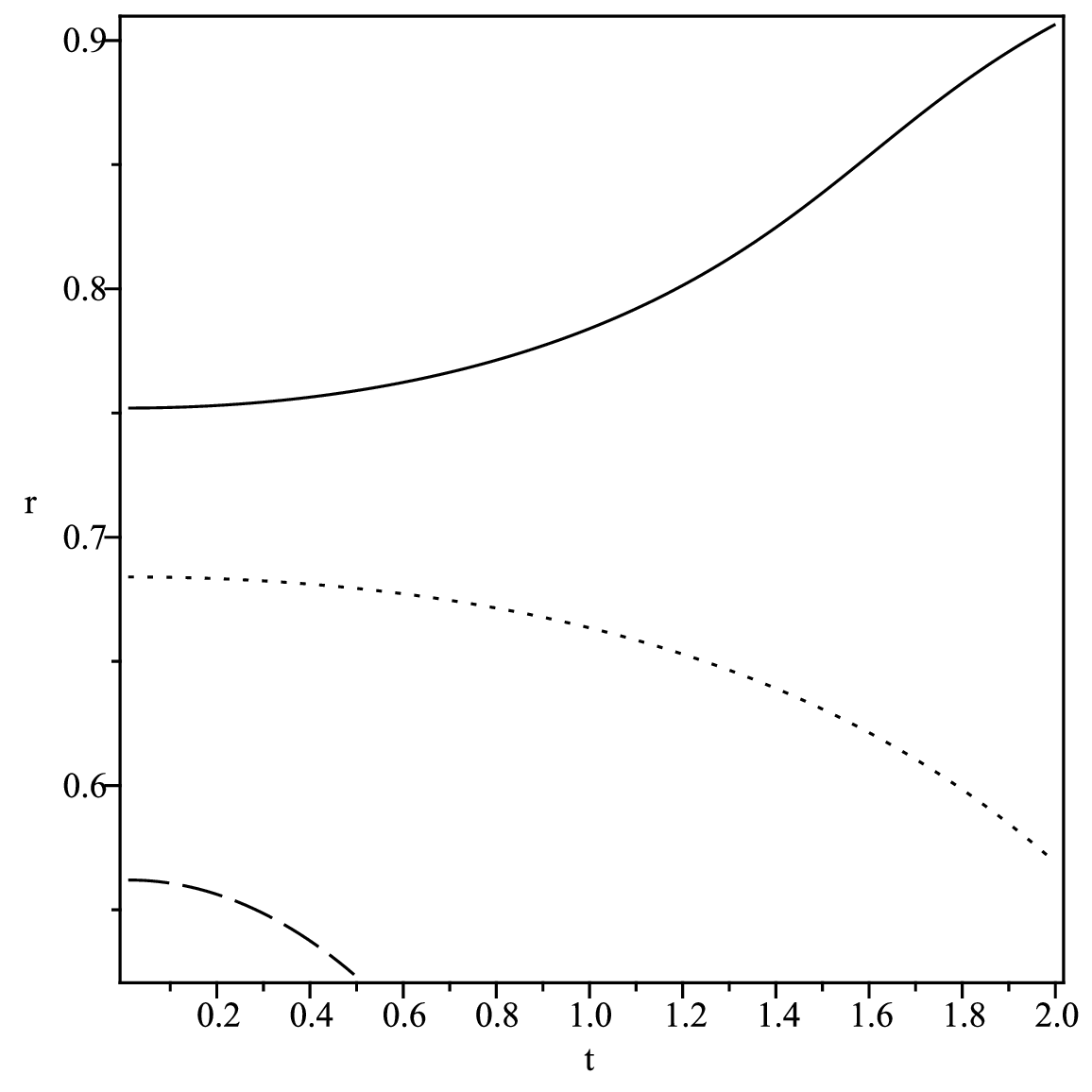}
\caption{Graph of the $r(t)$, radii of circular loops, in the BTZ
black hole with $M=1$ and $J=0.5$. The solid, dashed and dotted
curves drawn with initial radii $r(t_0)$=0.752, 0.562 and 0.684
respectively under assumption of $\dot{r}=0$ and $q=0$.}
\end{center}
\end{figure}

Now, we take $J$ as a constant and $M$ is different. So, we find
some constraint on the initial radii which give us the formation of
black holes.\\
\begin{center}
  \begin{tabular}{|c||c|}
    \hline
    $M$ &  $Critical        $ $r(t_0)$ \\ \hline\hline
    $For M<0$ & Any initial radii \\ \hline
    $M=0$ & Any initial radii \\ \hline
    1 & $r(t_0)>0.707$\\ \hline
    2 & $r(t_0)>0.369$\\ \hline
    3 & $r(t_0)>0.294$ \\ \hline
    10 & $r(t_0)>0.160$ \\ \hline
    \hline
  \end{tabular}
\end{center}

\hspace{2.50cm}Table (2). Results For $J=1$ and $q=0$, when $M$ is different.\\\\

As explained in the table 2, when the mass of the black hole is
$M\leq0$ there is no constraint on the initial radii of the cosmic
string loops for collapsing to form black holes. In that case there
is no chance for loops to survive in the space time and  variation
of $J$ has no affect to avoid the collapsing of the loops become
black hole. When the mass of the BTZ black hole is $M>0$ then the
loops for collapsing should satisfy constraint as $r(t_0)>\alpha$
and the values of $\alpha$ decrease with increasing of the BTZ mass.
We find that by increasing the mass of BTZ black hole the less loops
will stay because the initial radii is small, so there are more
loops satisfying this condition. Here, we must point out for the
spinless and $M=1$ BTZ black holes, we have de-Sitter space-time
which is solved by in the Ref. [33].

\begin{figure}[th]
\begin{center}
\includegraphics[scale=.4]{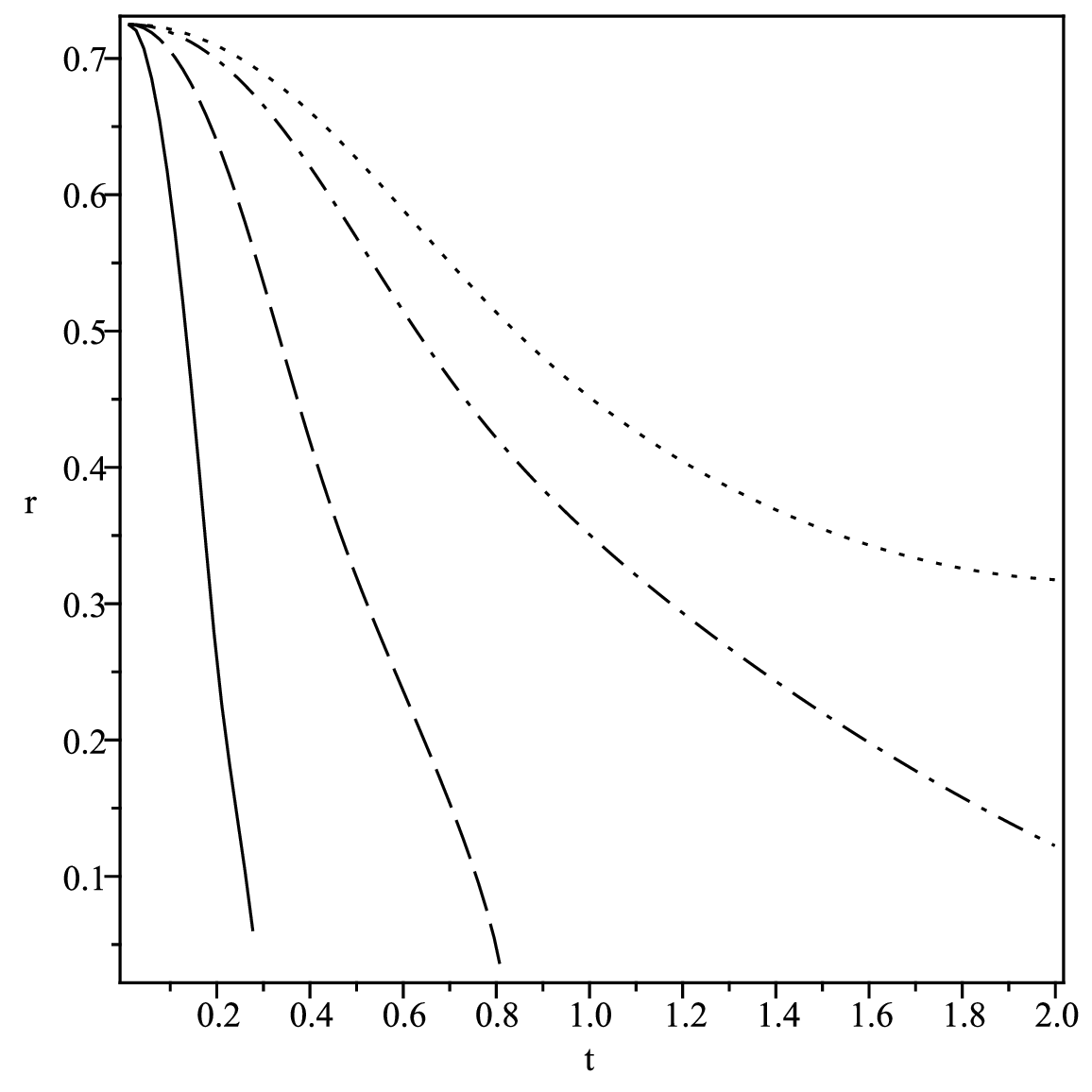}
\caption{Graph of the $r(t)$, radii of circular loops, in the BTZ
black hole with $J=1$. The solid, dashed, dash-dotted and dotted
curves drawn with $M=-3.8, -0.85, 0.002$, and 0.23 respectively with
initial values $r(t_0)=0.725, \dot{r}=0$ and $q=0$ .}
\end{center}
\end{figure}

In the special case, it is found for initial radii $r(t_0)<0.7070$
the loops will collapse to form black hole. But we showed that for
the BTZ black hole with angular momentum as $J\geq1$, the result is
clearly contrast with the results of the Ref. [33]. The physical
interpretation of this phenomena is simple, so the strings with
angular momentum have rotational motion and it yields to black hole
later. It means that the loops with condition like $r(t_0)>0.7070$
will collapse to form black hole, and also we note that smaller
loops avoid to become black hole. The observational results show
that there could exist a lot of cosmic string loops in our universe
[19, 25]. The final results of the cosmic string loops in the BTZ
background in the case of $J>1$ lead us to have lots of
loops.\\
Fig. 3 shows that if the magnitude of the BTZ black hole
mass decreases, then the rate of collapsing loops becomes faster. It means that massive loops are better candidate to forming black holes.\\
Next, we shall discuss about the cosmic strings with time-varying
tension. For the loops of cosmic string as the power of time
$\mu=\mu_0t^q$ we solve equation (9) numerically. As mentioned
above, the values of large $J$ can not cause collapsing loops to
black hole. Now, we consider $J=1$, $M=1$ and $\mu=\mu_0t^q$, and we
find that for all values of $q\geq 0$ our results are same as $q=0$.
It means that the loops satisfying $r(t_0)>0.707$ finally collapse
to black holes, and the value of $q>0$ can only has influence on the
time of collapsing (see the Fig. 4).

\begin{figure}[th]
\begin{center}
\includegraphics[scale=.38]{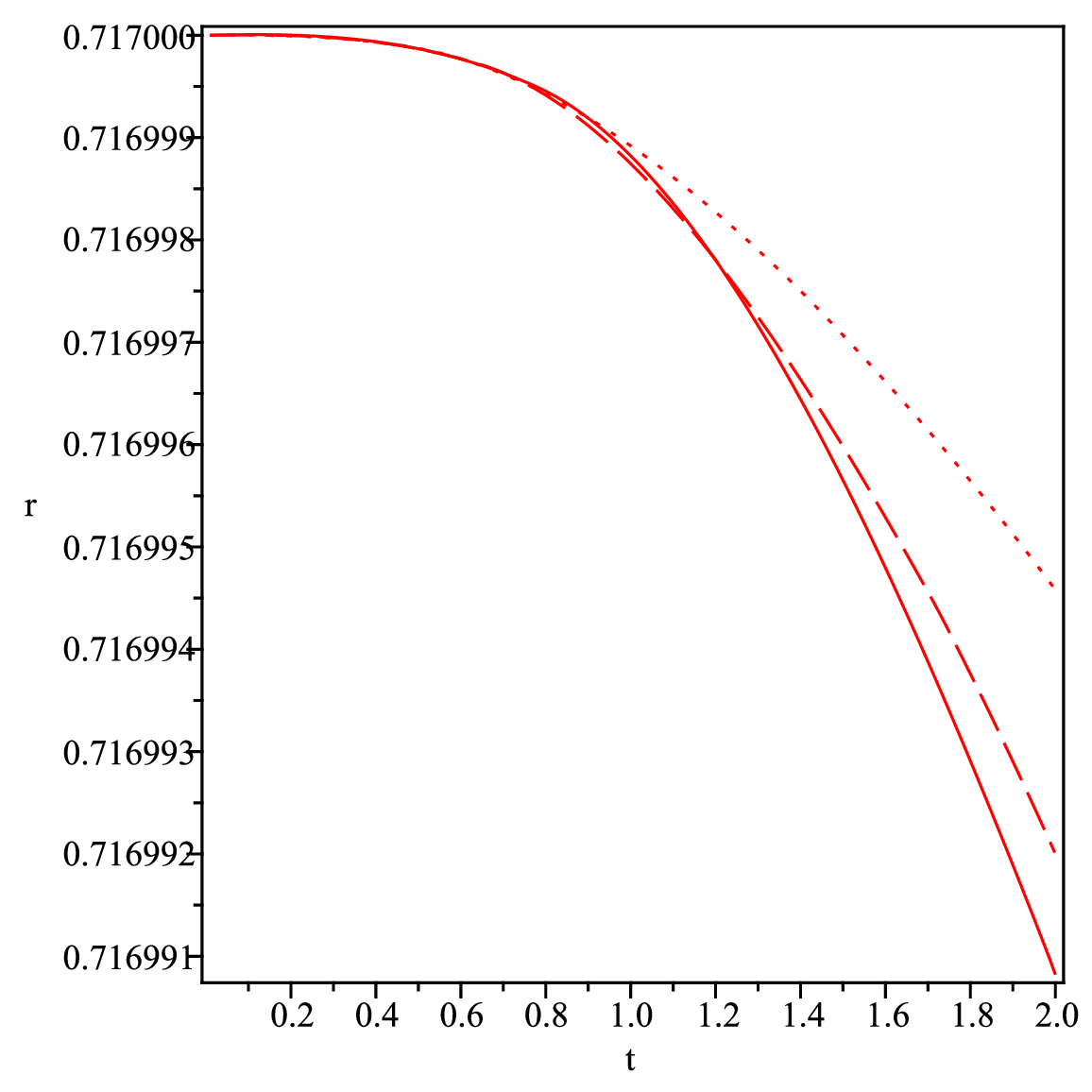}
\caption{The solid, dashed, and dotted curves of $r(t)$ represented
$q=1.5, 2.3$ and 4.5 respectively, where we assumed $M=1$, $J=1$,
$r(t_0) = 0.717$ and  $\dot{r}=0$.}
\end{center}
\end{figure}

The Fig. 4 shows that by increasing the values of $q$ the loops take
longer time to become black holes. But for $q<0$, when
$r(t_0)>0.707$, we can find limit of $q$ as, $-4.78<q<-0.65$ where
loops forming black holes, which is given in the Fig. 5. It means
that for $q\approx0$, where we have approximately time-independent
tension, loops can't form a black hole. Therefore, formation of
black hole is depend on time-dependence of string loops. Then by
increasing $|q|$ from 0.65 to 4.78 string loop can form a black hole
faster. As we know, for $q<0$, by increasing $|q|$ strings become
approximately tensionless ($\mu=\mu_0t^q$), so tensionless string
have no chance to form a black hole. Hence, there are two critical
values $q_{c_{1}}=0.65$ and $q_{c_{2}}=-4.78$, where string loops
with condition $q_{c_{1}}\geq q\geq q_{c_{2}}$ collapse to form
black hole. In another word $q>q_{c_{1}}$ yields to approximately
time-independent string and $q<q_{c_{2}}$ yields to approximately
tensionless string. We should note here that $q_{c_{2}}$ is larger
than the critical value of $q$ obtained from Minkowski of FRW
backgrounds.

\begin{figure}[th]
\begin{center}
\includegraphics[scale=.38]{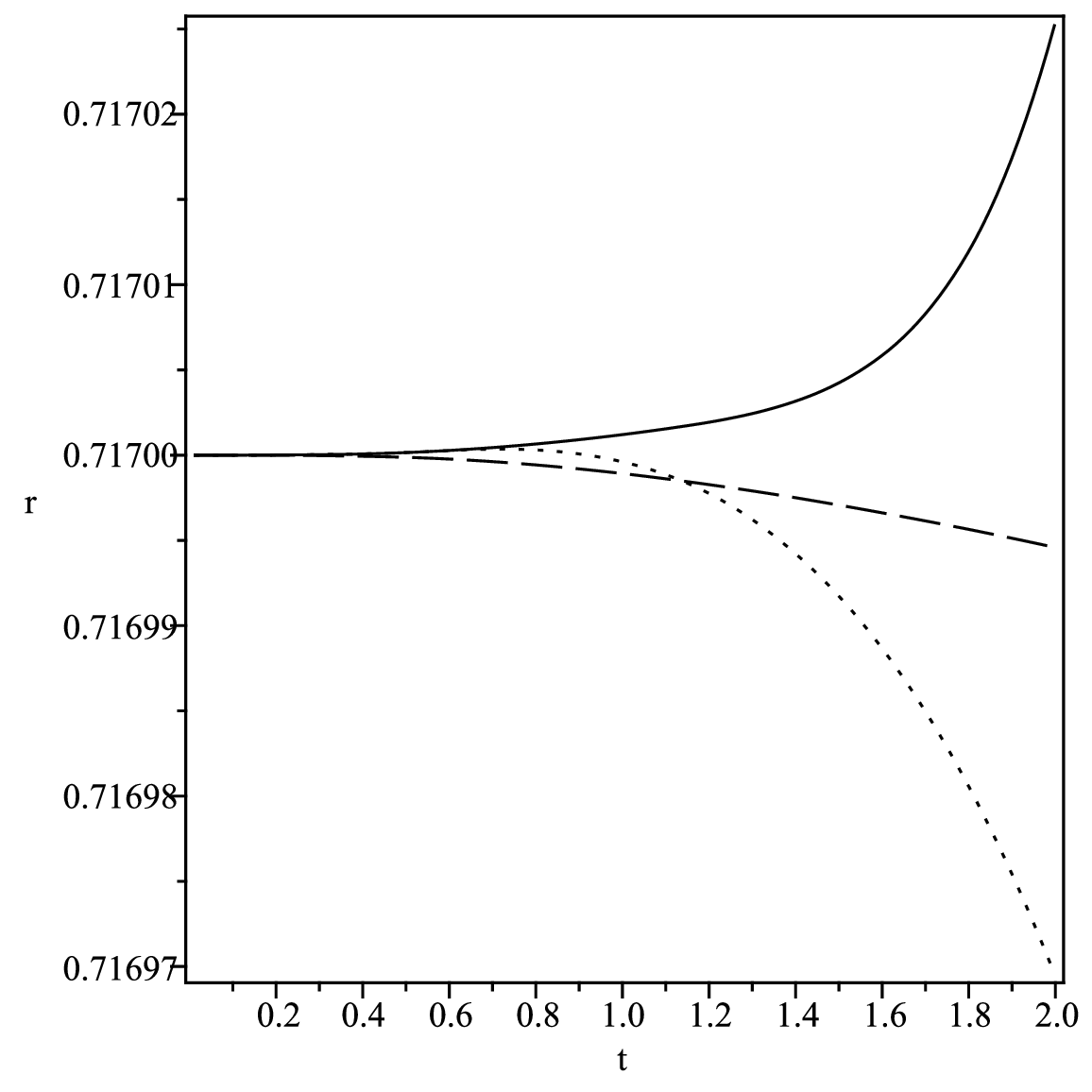}
\caption{The solid, dashed, and dotted curves of $r(t)$ as a
function of cosmic time represented $q=-7.9, -4.5$ and -2.3
respectively, where we assumed $M=1$, $J=1$, $r(t_0) = 0.717$ and
$\dot{r}=0$.}
\end{center}
\end{figure}

\section{Conclusion}
In this paper, we solved equation of a circular cosmic string loop
in the BTZ background, that evolving in the hypersurface with
$\theta = \frac{\pi}{2}$. By solving the equation (9) numerically,
we try to obtain various cases where cosmic string loop finally
change to form black holes. Then, we could obtain most of parameters
which have  effect on this event. First, we considered the tension
of loops as constant ($q=0$) for $J \neq 0$, and find a critical
value of initial radii of loops as $\alpha$, where if $ r(t_0) <
\alpha$ and $|J|<1$ loops will collapse to form black hole and for
$r(t_0)> \alpha$ and $|J|>1$ they will stay in universe. Interesting
case is $|J|>1$, so by increasing $J$ the critical value of radii
also increases. These  critical values are larger than the size of
universe. Thus, we can conclude for larger angular momentum, many of
cosmic string loops stay in universe and avoid to become black hole.
Also, we find that any change in the value of $M$ can lead to a new
critical initial radii, it means that by enlarging the value of the
BTZ black hole mass, the critical radii becomes larger and the
possibility of forming black holes become lesser. We can point out
for $M\leq 0$, without having any constraint, all the loops will
finally collapse to form black holes. Also we got to interesting
results for time-dependent tension of cosmic string loops, and found
that for all values of $q\geq 0$ any variation of $q$ has no effect
on the value of the critical radii. Then, the results are the same
as $q=0$. But by increasing the positive values of $q$ the loops
take longer time to become black holes (see Fig. 4). We have tried
to find the effect of $q<0$ on the evaluation of cosmic strings. In
that case, and by considering $J=1$, there is a limit for $q$ as
$-4.78<q<-0.65$ where loops become black holes. These results and
limited quantities may be useful to obtain loop number density
distribution which yield to compute various quantities describing
the cosmological impact of cosmic string loops today [46]. Such
studies may be lead to understand the mechanism of formation of
galaxies. This aim also may be verify
by consideration of cosmic strings as a possible source of geometric perturbations perpendicular to the string world-sheet [47].\\\\
{\bf Acknowledgments} It is pleasure to thank Hassan Firouzjahi for
reading the manuscript and giving physical comments and also
grammatical suggestions.

\end{document}